\documentclass[a4paper,11pt]{article}

\usepackage{jcappub}

\usepackage{slashed}
\usepackage{bm}
\usepackage{latexsym,amssymb,amsmath,float,url,mathrsfs}
\usepackage{latexsym}
\usepackage{epstopdf}
\usepackage{amsfonts}
\usepackage{amsmath}
\usepackage{amssymb}
\usepackage{comment}
\usepackage{subfigure}
\usepackage{natbib}
\usepackage{hyperref}
\usepackage{pifont}
\usepackage{blindtext}
\usepackage{adjustbox}
\usepackage{multirow}
\usepackage{tabularx}
\usepackage[table]{xcolor}
\usepackage{orcidlink}
\usepackage{tikz}
\usetikzlibrary{tikzmark}
\usepackage{amssymb}
\usepackage{bbold}
\usepackage{blkarray}
\usepackage{graphicx}
\usepackage{amsfonts}
\usepackage{soul}
\usepackage{amssymb}
\usepackage{amsmath}
\usepackage{cancel}
\usepackage{tcolorbox}
\usepackage{enumitem}

\usepackage{graphicx}
\usepackage{subcaption}

\usepackage{graphics,appendix,afterpage,makecell} 

\definecolor{oucrimsonred}{rgb}{0.6, 0.0, 0.0}
\definecolor{persianblue}{rgb}{0.11, 0.22, 0.73}
\definecolor{forestgreen}{rgb}{0.13,0.35,0.13}
\definecolor{lightgray}{rgb}{0.83, 0.83, 0.83}
 \hypersetup{colorlinks, citecolor=oucrimsonred, linkcolor=black, urlcolor=oucrimsonred}
\definecolor{cornellred}{rgb}{0.7, 0.11, 0.11}
\definecolor{navyblue}{rgb}{0.0, 0.0, 0.5}
\definecolor{amethyst}{rgb}{0.6, 0.4, 0.8}
\definecolor{yellow}{rgb}{1.0, 1.0, 0.0}
\definecolor{firebrick}{rgb}{0.7, 0.13, 0.13}
\definecolor{tangerineyellow}{rgb}{1.0, 0.8, 0.0}
\definecolor{deepfuchsia}{rgb}{0.76, 0.33, 0.76}
\definecolor{amber}{rgb}{1.0, 0.75, 0.0}
\definecolor{VioletRed4}{rgb}{0.55, 0.13, .32}
\definecolor{indiagreen}{rgb}{0.07, 0.53, 0.03}
\definecolor{VioletRed4}{rgb}{0.55, 0.13, .32}
\newcommand{\be}{\begin{equation}}
\newcommand{\ee}{\end{equation}}
\newcommand{\bea}{\begin{equation} \begin{aligned}}
\newcommand{\eea}{\end{aligned} \end{equation}}

\definecolor{oucrimsonred}{rgb}{0.6, 0.0, 0.0}
\newcommand\vertarrowbox[3][6ex]{%
  \begin{array}[t]{@{}c@{}} #2 \\
  \left\uparrow\vcenter{\hrule height #1}\right.\kern-\nulldelimiterspace\\
  \makebox[0pt]{\scriptsize#3}
  \end{array}%
}
\hypersetup{
     colorlinks   = true,
     citecolor    = violet,
     urlcolor     = violet,
     linkcolor    = violet}

\definecolor{verdechiaro}{rgb}{0.6,1,0.6}
\definecolor{giallochiaro}{rgb}{1,1,0.6}
\definecolor{bluscuro}{rgb}{0.15, 0.2, 0.9}
\definecolor{verdes}{rgb}{0.1, 0.5, 0.1}%
\definecolor{rossocorsa}{rgb}{0.83, 0.0, 0.0}
\definecolor{tangerineyellow}{rgb}{1.0, 0.8, 0.0}

\definecolor{americanrose}{rgb}{1.0, 0.01, 0.24}
\definecolor{cobalt}{rgb}{0.0, 0.28, 0.67}
\definecolor{brandeisblue}{rgb}{0.0, 0.44, 1.0}
\definecolor{mycolor}{rgb}{0.0, 0.0, 0.5}%navyblue
\definecolor{oxfordblue}{rgb}{0.0, 0.13, 0.28}
\definecolor{azure}{rgb}{0.0, 0.5, 1.0}
\definecolor{turquoiseblue}{rgb}{0.0, 1.0, 0.94}
\newtcolorbox{mynewbox}[1]{colback=white!5!white,colframe=azure!75!black,fonttitle=\bfseries,title=#1}
\newtcolorbox{mybox}{colback=mycolor!5!white,colframe=azure!75!black}
\newtcolorbox{mynamedbox}[1]{colback=mycolor!5!white,colframe=azure!75!black,title=#1}
\definecolor{venetianred}{rgb}{0.78, 0.03, 0.08}
\newtcolorbox{mynamedbox1}[1]{colback=venetianred!5!white,colframe=venetianred!80!black,title=#1}
\newtcolorbox{mynamedbox2}[1]{colback=azure!5!whia<tle=#1}

\newcommand{\td}{{\rm d}}

\definecolor{verdes}{rgb}{0.1, 0.5, 0.1}%
\definecolor{cornellred}{rgb}{0.7, 0.11, 0.11}

\newcommand{\nn}{\nonumber}

\definecolor{VioletRed4}{rgb}{0.55, 0.13, .32}

\title{Are Primordial Black Holes Truly Fine-Tuned?}
\author[a]{A.J.~Iovino\orcidlink{0000-0002-8531-5962},}
\author[b]{A.~Riotto\orcidlink{0000-0001-6948-0856}}

\affiliation[a]{New York University, Abu Dhabi, PO Box 129188 Saadiyat Island, Abu Dhabi, UAE}
\affiliation[b]{Department of Theoretical Physics and Gravitational Wave Science Center,  \\
24 quai E. Ansermet, CH-1211 Geneva 4, Switzerland}

\emailAdd{a.iovino@nyu.edu, antonio.riotto@unige.ch}

\abstract{Single-field inflationary models  which generate primordial black holes  through the  enhancement of the curvature primordial power at small scales are commonly criticized  and frequently  dismissed because they    require a large amount of fine-tuning  in the parameters setting the ultra slow-roll phase.  However, the  standardly adopted definition  of fine-tuning  has a clear drawback: the more the primordial black hole abundance is small and cosmologically harmless, the larger    the parameter space is fine-tuned.
A well-defined indicator of fine-tuning should assign large values to scenarios requiring significant parameter adjustment to reproduce the desired primordial black hole abundance, while yielding values of order unity in cases governed by standard sensitivity.
Motivated by such arguments, we use  the (modified version of)  Wilson's naturalness criterion for quantifying the fine-tuning and naturalness and  we show that the primordial black hole models are not technically unnatural.
}

\begin{document}
\maketitle
\flushbottom
\section{Introduction}
Primordial black holes (PBHs)~\cite{Carr:1974nx} have attracted increasing interest in recent years because of their potential roles in cosmology. They have been proposed as candidates for the dark matter (DM) component, as responsible for some of the observed gravitational-wave (GW) events, and as possible seeds of the supermassive black holes (SMBHs) in galactic nuclei. In scenarios where PBHs form from the gravitational collapse of large overdensities in the primordial curvature field~\cite{Ivanov:1994pa,Ivanov:1997ia}, obtaining a sizeable PBH abundance typically requires the curvature power spectrum to reach amplitudes of order $10^{-2}$ at the relevant scales. By contrast, on the large scales probed by the Cosmic Microwave Background (CMB), the inflationary spectrum is constrained to be of order $10^{-9}$~\cite{Planck:2018jri}. A mechanism is therefore needed to enhance the power spectrum at smaller scales.

Such an enhancement can be dynamically realised in single-field inflationary models that feature a phase of ultra-slow-roll (USR)~\cite{Ivanov:1994pa}. In these scenarios, after an initial standard slow-roll (SR) phase consistent with large-scale observations, the inflaton temporarily enters a phase called USR in which its motion nearly stalls (often due to a quasi-stationary inflection point in the potential), before resuming its evolution in a final SR phase that eventually ends inflation. All USR single-field models, however, involve a certain degree of fine-tuning of the potential parameters in order to satisfy cosmological constraints. Indeed, tuning the duration of the initial SR phase in order to achieve values of the inflationary observables, such as the scalar spectral index $n_s$ and the tensor to scala ratio $r$, compatible with the actual constraints\,\cite{Planck:2018jri,ACT:2025tim,SPT-3G:2025bzu}, and obtain the presence of a spectral peak high enough to trigger PBH production, can pose a notable challenge for model building.
In the community there is therefore a widespread feeling that the abundance of PBHs must be fine-tuned to produce a suitable abundance. 

Usually, to quantify the degree of fine-tuning of a given observable $X$, one  considers the logarithmic derivative of an observable $X$ with respect to the logarithm of any parameter $a$ on which $X$ depends, either explicitly or implicitly through intermediate quantities or convolutions:
\begin{equation}
\label{a}
    c(X,a) \equiv \left|\frac{a}{X}\,\frac{\partial X}{\partial a}\right|.
\end{equation}
This was assumed, for instance, in Ref.~\cite{Cole:2023wyx} for the abundance of PBHs.
However, the bare sensitivity $c$ to a parameter $a$ does not by itself constitute a measure of fine-tuning or naturalness. Although physical quantities depend sensitively on minute variations
of the fundamental parameters when there is fine-tuning, the latter is not
necessarily implied by the condition $c\gg 1$.  The expression  (\ref{a}) is really a measure of sensitivity, and sensitivity does
not automatically translate into fine-tuning\,\cite{Anderson:1994dz}. 

To illustrate this concept, we consider the same example of the proton mass discussed in Ref.~\cite{Anderson:1994dz}, which depends sensitively on small variations of the strong coupling constant at high energies $\mu$,

\be
\alpha_S(\mu)=\frac{\alpha_S(M_P)}{1-\beta_0\alpha_S(M_P)\ln(M_P/\mu)},
\ee
where $M_P$ is the Planck mass and  $\beta_0$ the one-loop QCD beta-function. Setting $\alpha_S(\mu=m_p)=1$ to define proton mass $m_p$ delivers 
\be
c(m_p,\alpha_S(M_P))\simeq \frac{1}{\beta_0}\frac{1}{\alpha_S(M_P)}\simeq  10^2.
\ee
The fact that $c(m_p,\alpha_S(M_P))$ is large reflects the extreme sensitivity of the proton mass to variations in $\alpha_S(M_P)$,  but the lightness of the proton
is a consequence of asymptotic freedom and the logarithmic running of the QCD gauge
coupling and not the result of unexplained cancellations. This sensitivity would persist independently of the proton’s numerical value, and therefore it is inappropriate to interpret a mass of ${\cal O}(1)$ GeV as evidence of fine-tuning.
Instead, the proton’s lightness is a direct consequence of asymptotic freedom and the logarithmic evolution of the QCD coupling, and hence poses no naturalness puzzle. 

Closely related issues were encountered in early discussions of naturalness within supersymmetric frameworks, where the pronounced dependence of superpartner masses on the top-quark mass~\cite{Barbieri:1988zs} artificially inflated estimates of fine-tuning by roughly an order of magnitude.

A reliable measure of fine-tuning must compare the sensitivity of a particular choice of parameters $c$ to a measure of the average, $\bar c$.  The problem with the fine-tuning definition (\ref{a}) applied to the abundance of the PBHs resides on the fact that it does give a lot of specific weight to regions  where the abundance is totally negligible, that is regions which are  totally natural from the model building point of view,  and pose no harm from
the cosmological point of view.  To see this point better, let us naively assume for the moment a gaussian distribution\footnote{A correct estimate of the abundance usinng the Press-Schechter formalism is presented in sec.2.2 and used in the rest of the work.} for the density contrast parameter $\delta$. As a consequence the PBH mass fraction reads\,\cite{Sasaki:2018dmp}  
\be
\beta\sim e^{-\delta_c^2/2A},
\ee
where $\delta_c\sim 0.5$ is the threshold for formation and $A$ is the variance of the density contrast.
Following the definition (\ref{a}), we will have

\be
c=\left|\frac{A}{\beta}\frac{\partial\beta}{\partial A}\right|=\frac{\delta_c^2}{2 A},
\ee
which is much  larger than unity  for  the CMB amplitudes $A\sim10^{-9}$ -- corresponding  to regions where $\beta\lll 1$  and the abundance of PBHs is totally negligible --  and decreases for larger values where the PBH abundance is relevant. In other words, the definition (\ref{a}) delivers a misleading large fine-tuning for values of the amplitude $A$ which are in fact perfectly admittable from the cosmological point of view. Similarly to the case of the proton mass, a large value of $c$, and therefore a large fine-tuning,  is only an accidental consequence of the fact that required value of $A$ is smaller than unity, but this does not imply  that the PBH abundance is really fine-tuned in a technical sense.

Inspired by the work of  Refs.~\cite{Anderson:1994dz,Anderson:1994tr,Anderson:1996ew}, we therefore adopt  a different measure of  fine-tuning 
\begin{equation}\label{eq:gamma}
    \gamma \equiv c/\bar{c},
\end{equation}
where
\begin{equation}
\label{criterion}
    \bar{c} \equiv 
    \frac{\int da \, c(X,a) }{\int da }.
\end{equation}
This definition is equivalent to Wilson’s naturalness criterion~\cite{Susskind:1978ms}, according to which the observable properties of a system should not be unusually unstable under small variations of the fundamental parameters of the theory. As a consequence, an ideally natural model satisfies $\gamma \simeq 1$, whereas strongly fine-tuned models correspond to $\gamma \gg 1$ \textcolor{magenta} {or $\gamma\ll 1$}.

The distinction between $c$ and $\gamma$ can be understood as follows. The parameter $c$ quantifies sensitivity, namely how much a given physical observable (such as the amplitude or the PBH abundance) changes under a small variation of an underlying parameter, which does not necessarily have a direct physical interpretation (for instance, a generic coefficient in the potential). By contrast, $\gamma$ is designed to capture how this sensitivity compares to its typical value across parameter space. In particular, it provides a measure of whether the sensitivity along a given direction is representative or instead atypically large or small when compared to an average over directions.

To clarify this point, it is useful to adopt a geometric perspective and consider a multidimensional parameter space. A displacement in a given direction corresponds to a variation of the model parameters, and the magnitude of this displacement required to produce a fixed change in the physical observable is precisely what $c$ measures. In this sense, $c$ characterizes the local sensitivity along a given direction in parameter space.

On the other hand, $\gamma$, by construction, involves a normalization with respect to an average over directions. As a result, $\gamma \sim \mathcal{O}(1)$ indicates that all directions in parameter space are equivalent, i.e.\ there is no preferred direction. Conversely, if $\gamma \gg 1$ along a given direction, this signals that the model exhibits a strong directional dependence, meaning that certain parameter variations are significantly more effective than others. This can be interpreted as a lack of naturalness, since a priori equivalent directions in parameter space do not lead to comparable physical effects.

In summary, while $c$ measures absolute sensitivity along a given direction, $\gamma$ quantifies how this sensitivity compares to the typical behavior across parameter space.

The rest of the paper is dedicated to show that with such a definition the PBH abundance is not technically fine-tuned. The paper is organized as followed. We examine in Sec.~\ref{sec:Anal} a set of benchmark USR models and in Sec. \ref{sec:Results} their corresponding degrees of fine-tuning. We conclude in Sec.~\ref{sec:Conc}. Throughout this work we adopt units $c = M_{\rm Pl} = \hbar = 1$.

\section{Benchmark models}\label{sec:Anal}
We consider a real scalar field coupled to gravity in the so-called Jordan frame:
\begin{equation}
\mathcal{S}=\int d^4 x \sqrt{-g}\left(-\frac{1}{2}\left(1+\xi \phi^2\right) R+\frac{1}{2} g_{\mu \nu} \partial^\mu \phi \partial^\nu \phi-V(\phi)\right).
\end{equation}
The non minimal coupling between $\phi$ and $R$ can be recast into a non-canonical kinetic term for the scalar field by performing a Weyl transformation of the metric\footnote{For simplicity, we do not consider non-minimal terms of dimension greater than four although they may be present in the context of a generic effective field theory.}:
\begin{equation}
\tilde{g}_{\mu \nu}(x)=\Omega^2[\phi(x)] g_{\mu \nu}(x) \quad \text { with } \quad \Omega^2=1+\xi \phi^2.
\end{equation}
This leads to
\begin{equation}
\mathcal{S}=\int d^4 x \sqrt{-\tilde{g}}\left(-\frac{1}{2} \tilde{R}+\frac{1}{2} K(\phi) \tilde{g}_{\mu \nu} \partial^\mu \phi \partial^\nu \phi-U(\phi)\right).
\end{equation}
where
\begin{equation}
K=\frac{1}{\Omega^2}+\frac{3 M_P^2}{2}\left(\frac{d \log \Omega^2}{d \phi}\right)^2 \quad \text { and } \quad U=\frac{V}{\Omega^4}.
\end{equation}
The kinetic term of the scalar field can be canonically normalized with the following field redefinition:
\begin{equation}
\Omega^2 \frac{d \bar{\phi}}{d \phi}=\sqrt{\Omega^2+\frac{3 M_P^2}{2}\left(\frac{d \Omega^2}{d \phi}\right)^2}.
\end{equation}
Then the action reads
\begin{equation}
\mathcal{S}=\int d^4 x \sqrt{-\tilde{g}}\left(-\frac{1}{2} \tilde{R}+\frac{1}{2} \tilde{g}_{\mu \nu} \partial^\mu \bar{\phi} \partial^\nu \bar{\phi}-U(\bar{\phi})\right).
\end{equation}
This is the form of the action that we will use in later sections of the paper to compute the inflationary dynamics and the abundance of PBHs from the potentials in Sec.\ref{PS} and Sec.\,\ref{Abu}.
In the non minimally coupled scenarios, we simply have $\phi=\bar{\phi}$ and $V(\phi)=U(\bar{\phi})$.

In our analysis, trying to mimic the previous literature, we focus on the following potentials:
\begin{enumerate}[label=(\(\mathcal{\Alph*}\))]
\item \textbf{\textcolor{rossocorsa}{Toy model with an artificial dip or bump}~\cite{Atal:2019cdz,Mishra:2019pzq,Inomata:2021tpx,Bhatt:2022mmn}:} \\
One of the simplest ways to construct a successful single field model is to consider a generic minimally coupled inflaton potential that satisfies CMB constraints on large scales and to introduce by hand a localized feature, such as a bump or a dip, in order to generate an USR phase at smaller scales. Although ad hoc in nature, such modifications provide a simple and effective framework for building toy models that enhance the primordial power spectrum and lead to PBH formation.  

In our case, we focus on Starobinsky model of inflation with a bump described by
\begin{align}
V_{\textrm{Staro+dip}}(\phi) \equiv 
V_{\textrm{Staro}}(\phi)\left[
1-A_s\cosh^{-2}\left(\frac{\phi-\phi_0}{\sigma}\right)
\right].\label{eq:StaroDip}   
\end{align}
with
\begin{equation}
   V_{\textrm{Staro}}(\phi) = 
\frac{3M^4}{4}
\left[1 - 
\exp\left(
-\sqrt{\frac{2}{3}}
\phi
\right)
\right]^2,
\end{equation}
where the base potential $V_{\rm Staro}$ corresponds to the usual Starobinsky inflationary model~\cite{Starobinsky:1980te}, while the hyperbolic  function—characterised by its amplitude $A_s$, position $\phi_0$, and width $\sigma$—acts as a speed-breaker for the inflaton, allowing for the realization of a USR phase.

In all the potentials considered in this work, $M$ is a fundamental mass scale and the inflaton field $\phi$ is expressed in unit of Planck mass.

\item \textbf{\textcolor{verdes}{Minimally coupled polynomial model}~\cite{Hertzberg:2017dkh,Allegrini:2024ooy,Allegrini:2025jha}:} \\
Finite-order polynomial potentials naturally provide a mechanism to decelerate the inflaton background and generate a USR phase through the presence of one or more inflection points. Arranging an inflection point in field space that sufficiently enhances the primordial power spectrum can be achieved through a local cubic or higher-order polynomial expansion. Since a sufficiently large scalar spectral index requires the inclusion of terms of at least dimension six in minimally coupled polynomial potentials~\cite{Ballesteros:2020qam}, we focus on the following model~\cite{Allegrini:2024ooy}:
\begin{equation}\label{eq:MC}
    V(\phi) = b_4 M^4
    \left(
        \bar{b}_2 \phi^2
        + \bar{b}_3 \phi^3
        + \phi^4
        + \bar{b}_5 \phi^5
        + \bar{b}_6 \phi^6
    \right).
\end{equation}
where $\bar{b}_i\equiv b_i/b_4$.
\item \textbf{\textcolor{brandeisblue}{Non-minimally coupled polynomial model}~\cite{Kannike:2017bxn,Garcia-Bellido:2017mdw,Ballesteros:2017fsr,Germani:2017bcs,Frosina:2023nxu}:} \\
The presence of a non-minimal coupling to gravity generally flattens the inflaton potential at large field values, leading to a larger scalar spectral index compared to the minimally coupled case~\cite{Ballesteros:2020qam}. Following the existing literature, we consider as before a polynomial of dimension 6 with a non-minimal coupiling $\xi$~\cite{Ballesteros:2020qam}. After a Weyl transformation of the metric the potential is described by
\begin{equation}\label{eq:NMC}
    U(\phi) = \frac{a_4 M^2}{(1+\xi\phi^2)^2}
    \left(
        \bar{a}_2 \phi^2
        + \bar{a}_3 \phi^3
        + \phi^4
        + \bar{a}_5 \phi^5
        + \bar{a}_6 \phi^6
    \right).
\end{equation}
where again $\bar{a}_i\equiv a_i/a_4$.
\end{enumerate}

\subsection{Computing the Power Spectra}\label{PS}

The Hubble-flow parameters $\epsilon_{i}$ (for $i\geqslant 1$) are defined by the recursive relation
\begin{align}
\epsilon_{i} \equiv \frac{\dot{\epsilon}_{i-1}}{H\epsilon_{i-1}}\,,~~~~~\textrm{with:}~~~
\epsilon_0 \equiv \frac{1}{H}\,.\label{eq:HubblePar1}
\end{align}
where in this expression, $a=a(t)$ is the scale factor,
the Hubble expansion rate is defined by $H \equiv \dot{a}/a$ where the overdot denotes a derivative with respect to cosmic time, that is, $\dot{a}=da/dt$. As customary, we simply indicate as $\epsilon$ the first Hubble parameter, $\epsilon \equiv \epsilon_1 = -\dot{H}/H^2$. 
Instead of the second Hubble parameter $\epsilon_2$, sometimes it is useful to introduce the Hubble parameter $\eta$ 
defined by
\begin{align}
\eta \equiv - \frac{\ddot{H}}{2H\dot{H}} 
= \epsilon - 
\frac{1}{2}\frac{d\log\epsilon}{dN}\,,~~~~
\textrm{with:}~~~\epsilon_2 = 2\epsilon - 2\eta\,.\label{eq:HubblePar2}
\end{align}
Using the number of $e$-folds as time variable,  the redefined inflaton equation of motion reads
\begin{align}
\frac{d^2\bar{\phi}}{dN^2} + \left[3 - \frac{1}{2}\left(\frac{d\bar{\phi}}{dN}\right)^2\right]
\left[\frac{d\bar{\phi}}{dN} + \frac{d\log U(\bar{\phi})}{d\bar{\phi}}
\right] = 0\,,\label{eq:EoM}
\end{align}
with the Hubble rate that is related to the modified inflaton potential by means of the Friedmann equation
\begin{align}
(3-\epsilon)H^2 = U(\bar{\phi}) \quad \textrm{and} \quad \epsilon=(d\bar{\phi}/dN)^2/2\,.\label{eq:HEvo}
\end{align}
As in Ref.~\cite{Ballesteros:2020qam}, we define the USR phase by the condition $\epsilon_2 < -3$ instead of the narrower definition $\epsilon_2 = - 6$.

In order to compute the scalar power spectrum of curvature perturbations we need to  solve the Mukhanov-Sasaki (MS) equation \cite{Sasaki:1986hm,Mukhanov:1988jd}
\begin{align}\label{eq:M-S}
\frac{d^2 u_k}{dN^2} &+ (1-\epsilon)\frac{du_k}{dN} + 
\left[
\frac{k^2}{(aH)^2} + (1+\epsilon-\eta)(\eta - 2) - \frac{d}{dN}(\epsilon - \eta)
\right]u_k = 0\,,
\end{align}
with sub-horizon Bunch-Davies initial conditions\,\cite{Bunch:1978yq} at $N \ll N_k$, where $N_k$ indicates the horizon crossing time for the mode $k$, that is the time at which we have $k = a(N_k)H(N_k)$. 
The scalar power spectrum of curvature perturbations can be written as
\begin{align}\label{eq:PS}
P_{\zeta}(k) = \frac{k^{3}}{2\pi^{2}}
\left|\frac{u_k(N)}{a(N)\,\bar{\phi}'(N)}\right|^{2}_{\,N > N_{\rm F}(k)}\, .
\end{align}
This quantity is effectively time–independent, since Eq.~\eqref{eq:PS} is to be evaluated only after the mode reaches the freeze-out time $N_{\rm F}(k)$, i.e.\ when the combination $|u_k/a\bar{\phi}'|$ has settled to a constant value that remains conserved until horizon re-entry. We define the freeze-out time as
\begin{equation}
N_{\rm F}(k) \equiv \max\{N_k,\, N_{\rm end}\},
\end{equation}
where $N_{\rm end}$ marks the end of the USR phase. Modes with $N_k < N_{\rm end}$ exit the horizon before the end of USR and do not freeze immediately: although they are already super-horizon, they later experience the negative-friction regime. Their contribution to Eq.~\eqref{eq:PS} must therefore be evaluated at any time $N > N_{\rm end} > N_k$. Conversely, modes with $N_k > N_{\rm end}$ leave the horizon only after the USR phase has finished and thus freeze as soon as they become super-horizon. For these modes, Eq.~\eqref{eq:PS} may be evaluated at any time $N > N_k > N_{\rm end}$.
\subsection{Computing the PBH abundance}\label{Abu}
The fraction of energy density $\beta_k(M_{\rm PBH}) \td  \ln M_{\rm PBH}$ collapsing into PBHs can be estimated as 
\be\label{eq:betak}
    \beta_k(M_{\rm PBH})
    = \int_{\mathcal{C}_{\rm th}} \! \td\mathcal{C} \, P_k(\mathcal{C}) \frac{M_{\rm PBH}}{M_k}  \delta\left[ \ln\frac{M_{\rm PBH}}{M_{\rm PBH}(\mathcal{C})} \right]\!,
\ee
where $P_k(\mathcal{C})$ denotes the probability a black hole will form in the Hubble patch. The PBH mass function can be obtained directly from the collapse fraction:
\be\label{eq:df_PBH}
      \!\frac{\td f_{\rm PBH}}{\td \ln M_{\rm PBH}}
= \frac{1}{\Omega_{\rm DM}}\int \frac{\td M_k}{M_k} \, \beta_k(M_{\rm PBH} ) \left(\frac{M_{\rm eq}}{M_k}\right)^{1/2} \!\!,
\ee
where $M_{\rm eq} \approx 2.8\times 10^{17}\,\,M_{\odot}$ is the horizon mass at the time of matter-radiation equality and $\Omega_{\rm  DM} = 0.12h^{-2}$ is the cold dark matter density~\cite{Planck:2018jri}.
The total PBH abundance is given by
\begin{equation}
   f_{\rm PBH} 
   = \int \frac{\td M_{\rm PBH}}{M_{\rm PBH}} \frac{\td f_{\rm PBH}}{\td \ln M_{\rm PBH}}\,.
\end{equation}
The PBH mass function depends on the formalism adopted to describe PBH formation. In this work, we follow the approach based on threshold statistics of the compaction function, in which the probability of PBH formation is estimated from the statistical properties\footnote{Since the level of primordial non-Gaussianity generated in USR models is typically small~\cite{Atal:2018neu,Franciolini:2022pav,Firouzjahi:2023xke,Frosina:2023nxu}, we safely neglect its impact on the PBH abundance~\cite{Young:2022phe,Ferrante:2022mui,Gow:2022jfb,Ianniccari:2024bkh}.} of the compaction function $\mathcal{C}$~\cite{Ferrante:2022mui,Gow:2022jfb}, which is generically defined as twice the local mass excess over the areal radius.\footnote{An alternative formalism is based on peak theory~\cite{Young:2014ana,Yoo:2018kvb,Yoo:2019pma,Franciolini:2022tfm,Young:2020xmk,Musco:2023dak}. Since the two approaches yield similar PBH abundances, differing mainly by an overall shift of a few orders of magnitude, we expect the degree of fine-tuning to be largely insensitive to the specific formalism adopted.}.
The resulting mass function is given by\,\cite{Young:2019yug,DeLuca:2019qsy}
\begin{equation}
\begin{aligned}
f_{\rm PBH}(M_{\rm PBH}) = \frac{1}{\Omega_{\rm DM}}\int_{M_H^{\rm min}}^{\infty}\frac{\td M_H}{M_H}
\left(
\frac{M_{\rm eq}}{M_H}
\right)^{1/2}\left(\frac{M_{\rm PBH}}{\mathcal{K}M_H}
\right)^{1/\gamma}\left(
\frac{M_{\rm PBH}}{\gamma M_H}
\right) 
\frac{
\textrm{Exp}\Bigg[-\frac{8\left(1 - 
\sqrt{\Lambda}\right)^2}{9\sigma_c^2(M_H)}\Bigg]
}{
\sqrt{2\pi}\sigma_c(M_H)
\Lambda^{1/2}}
\,,
\end{aligned}
\end{equation}
where 
\begin{equation}
\Lambda=1 - \left({\cal C}_{\rm th} - \frac{3\left(M_{\rm PBH}/(\mathcal{K}M_H)\right)^{1/\gamma}}{2}\right)
\end{equation}
and the lower limit of integration follows from the condition $\Lambda > 0$.

The variance can be computed as
\begin{equation}
    \sigma_c^2(M_H) = \frac{16}{81}
\int_{0}^{\infty}\frac{\td k}{k}(k r_m)^4W(k r_m)^2 P_{\zeta}^{\mathcal{T}}(k,r_m)\label{eq:Sigma2}
\end{equation}
with $P_\zeta^{\mathcal{T}}={\mathcal{T}}^2\left(k, r_m\right) P_\zeta(k)$. We have defined $W\left(k, r_m\right)$ and ${\mathcal{T}}\left(k, r_m\right)$ as the top-hat window function and the radiation transfer function\,\cite{Young:2022phe}.
Although the parameters $\gamma$, $r_m$, $\mathcal{K}$ and $C_{\rm th}$ depend on the shape of the power spectrum\,\cite{Musco:2020jjb,Byrnes:2018clq,Musco:2023dak,Ianniccari:2024ltb} and on the thermal history of the universe\,\cite{Musco:2023dak}. For simplicity in this work we assume a radiation dominated universe and we take $\gamma=0.36$, $C_{\rm th}=0.55$, $r_m=3/k_{\rm peak}$, $\mathcal{K}=5$.
\begin{figure}[!t]
    \centering
    \includegraphics[width=0.8\textwidth]{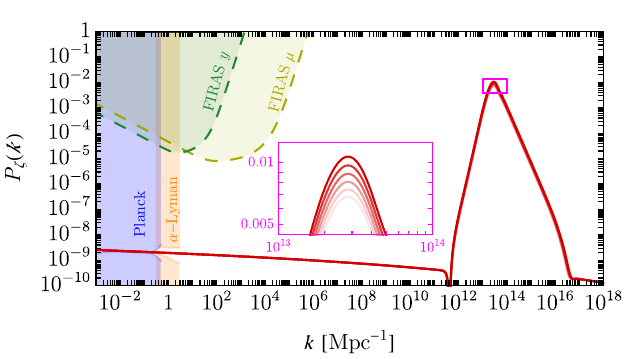}
    \caption{  \it
   On the main panel, we show power spectra of curvature perturbations over the entire range of scales covered by the inflationary dynamics generate in the Toy model with an artificial dip, Eq.\eqref{eq:StaroDip} tuning the parameter $A_s$ while we fix the others parameters.  We also plot the region excluded by CMB anisotropy measurements, Ref.\,\cite{Planck:2018jri}, the FIRAS bound on CMB spectral distortions, Ref.\,\cite{Chluba:2012we} (see also Ref.\,\cite{Jeong:2014gna,Iovino:2024tyg}) and the bound obtained from Lyman-$\alpha$ forest data \cite{Bird:2010mp}. The magenta inset shows the same Power spectra of curvature perturbations as a function of comoving wavenumbers, zoomed in on the region of scales relevant for PBH productions. 
    }
\label{Fig:PS}
\end{figure}
\section{Results}
\label{sec:Results}
In our analysis, we generate curvature power spectra whose dominant peak is located at 
$k_{\rm peak} \simeq 10^{13}\,\mathrm{Mpc}^{-1}$, corresponding to an average PBH mass of 
$\sim 10^{-14}\,M_\odot$, a mass range in which PBHs can account for the entirety of the dark matter abundance~\cite{Carr:2026hot}. 
All configurations considered in this work satisfy $n_s \simeq 0.96$ and $r \lesssim 0.06$, in agreement with the Planck constraints~\cite{Planck:2018jri}.\footnote{Recently, ACT data~\cite{ACT:2025tim} combined with Planck and DESI have shown a notable shift towards a nearly scale-invariant spectrum, yielding $n_s = 0.9743 \pm 0.0034$, which differs from the previous Planck constraint by approximately $2\sigma$. Shortly thereafter, the SPT collaboration~\cite{SPT-3G:2025bzu} reported a measurement fully consistent with the Planck result, $n_s = 0.9684 \pm 0.0030$. Since resolving this tension lies beyond the scope of this work, we adopt the results of the well-established Planck analysis~\cite{Planck:2018vyg}.}

In order to obtain a phenomenologically viable PBH abundance, we tune the parameters of the inflationary potential one at a time, keeping the others fixed, such that the amplitude of the power spectrum lies in the range 
$A \sim [0.007,\,0.01]$, which in turn yields a relevant PBH fraction 
$f_{\rm PBH} \sim [10^{-6},\,1]$.\footnote{The precise value of the PBH abundance as a function of the amplitude depends on the specific inflationary model under consideration, since different models produce mass functions with slightly different shapes due to variations in the width of the power spectrum, leading to moderately different abundances.}
Examples of the resulting power spectra are shown in Fig.~\ref{Fig:PS}.

\begin{figure}[t!]
    \centering

    % --------- Primo plot ----------
    \begin{minipage}{\textwidth}
        \centering
        \includegraphics[width=0.99\textwidth]{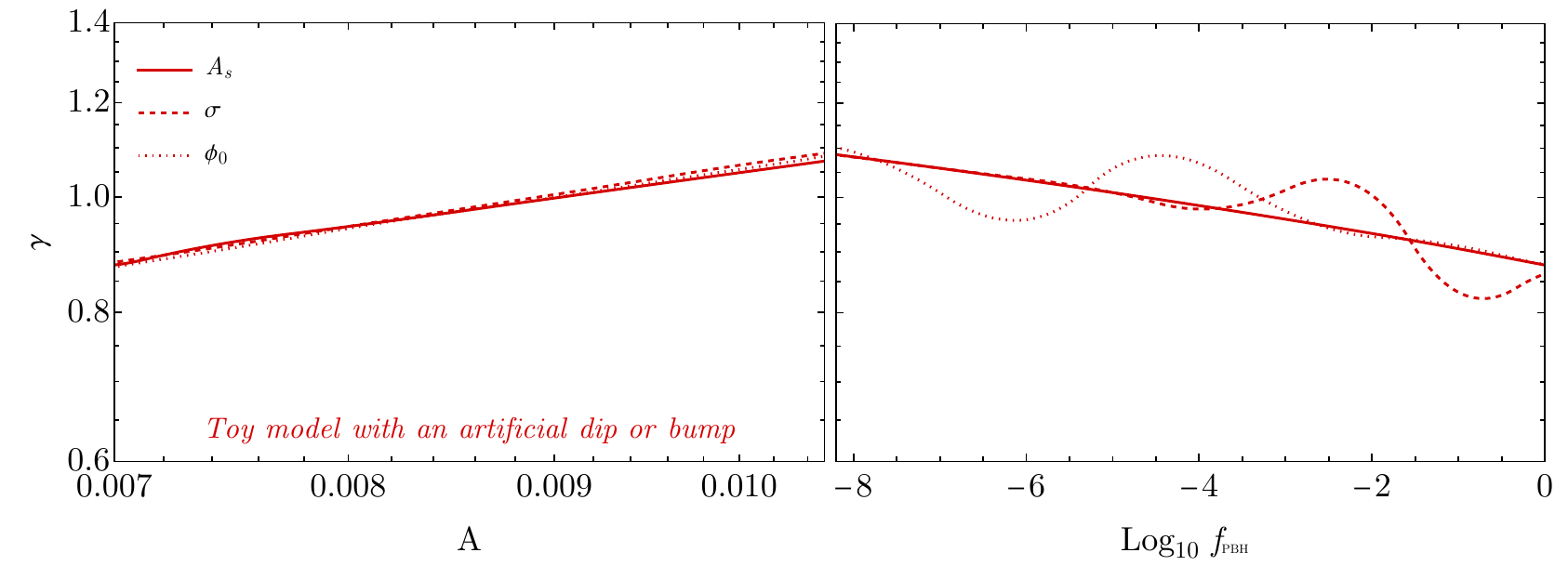}
        \label{fig:plot1}
    \end{minipage}

    % --------- Secondo plot ----------
    \begin{minipage}{\textwidth}
        \centering
        \includegraphics[width=0.99\textwidth]{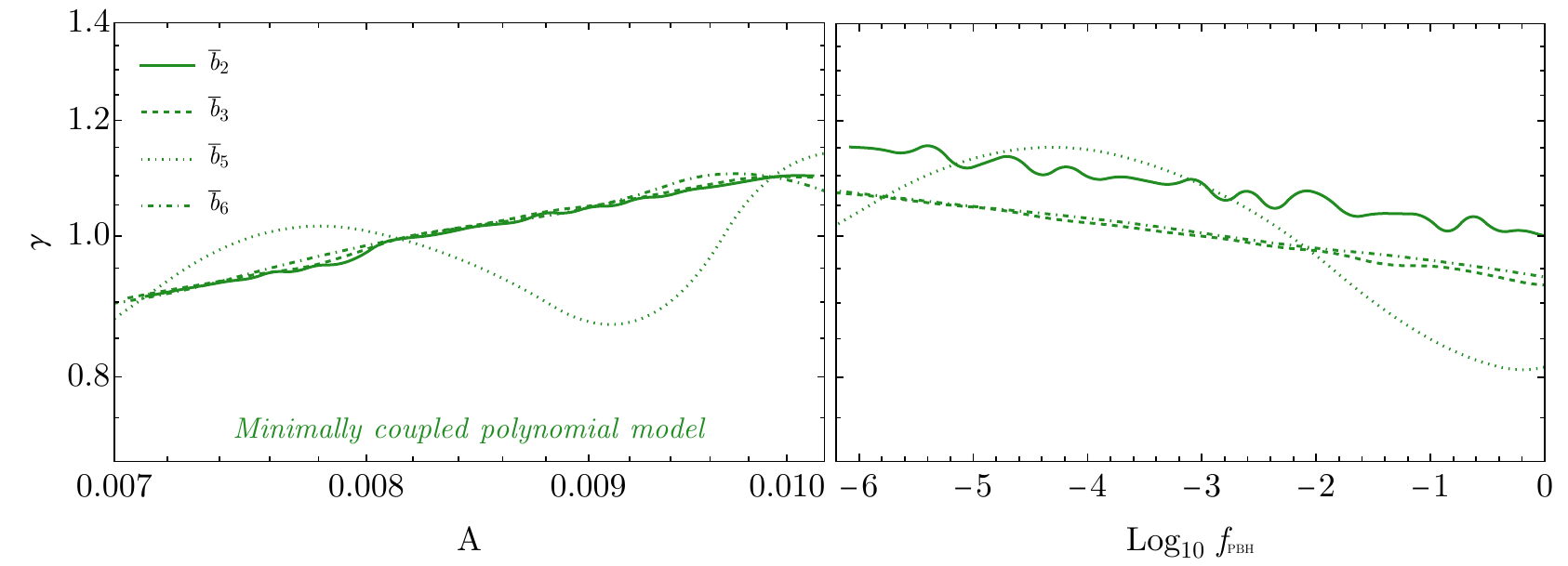}
        \label{fig:plot2}
    \end{minipage}

    % --------- Terzo plot ----------
    \begin{minipage}{\textwidth}
        \centering
        \includegraphics[width=0.99\textwidth]{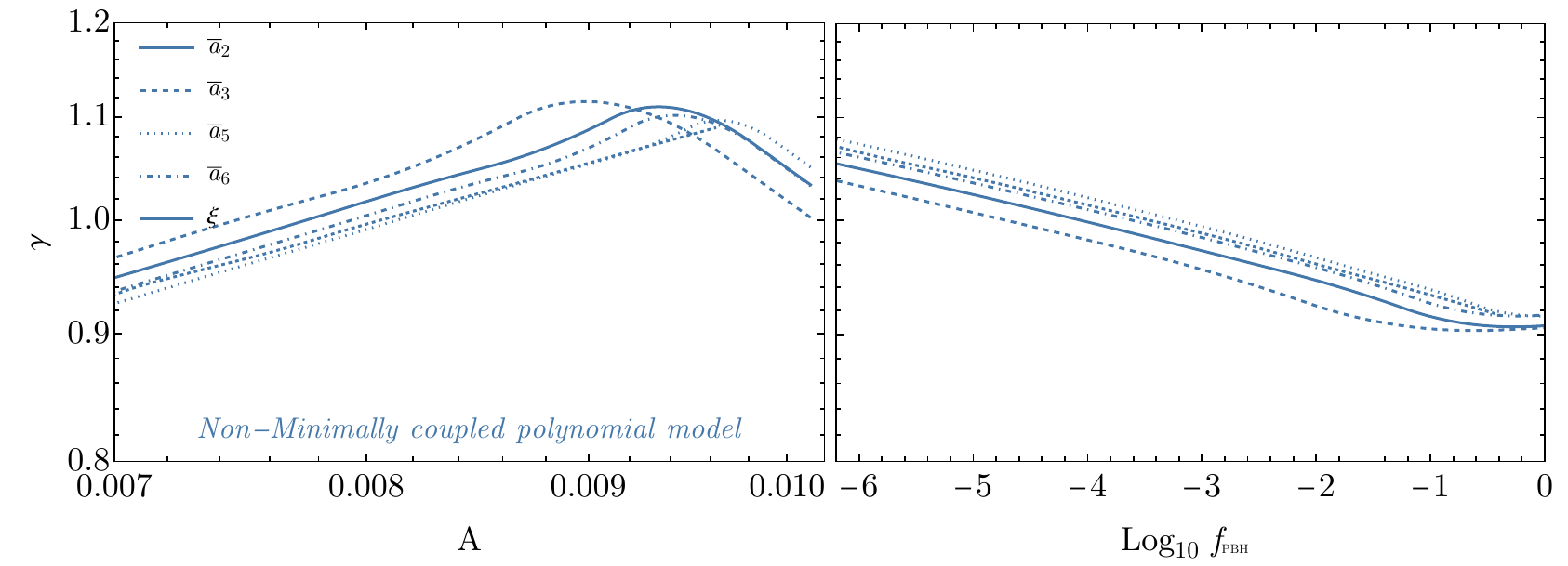}
        \label{fig:plot3}
    \end{minipage}

    % --------- Caption comune ----------
    \caption{
    \it Evolution of the naturalness parameter $\gamma$ for the toy model with an artificial dip, Eq.\eqref{eq:StaroDip}, (top panel), minimally coupled polynomial model, Eq.\eqref{eq:MC}, (middle panel) and non-minimally coupled polynomial model, Eq.\eqref{eq:NMC}, (bottom panel) as function of the amplitude of the main peak $A$ (left panels) and of the corresponding PBH abundance $f_{\rm PBH}$ (right panels).
    }

    \label{fig:fine-tuning}
\end{figure}

For each class of inflationary potential $V(\phi)$ considered in this work, we quantify the degree of fine-tuning by computing the fine-tuning parameter $c$ and the Wilson's naturalness parameter $\gamma$, varying a subset of the fundamental parameters characterising the potential. 
\\For each set of models we find large values of $c$. Indeed tuning the amplitude $A$ we find
\begin{align}
    \textrm{Toy model with an artificial dip} &\quad c(A_s,\sigma,\phi_0)\sim (7\cdot10^3,5\cdot10^3,3\cdot10^4)\nn \\
    \textrm{Minimally coupled polynomial} &\quad c(\bar{b}_2,\bar{b}_3,\bar{b}_5,\bar{b}_6)\sim (5\cdot10^7,10^8,4\cdot10^7,4\cdot10^6)\nn \\
    \textrm{Non-Minimally coupled polynomial} &\quad c(\bar{a}_2,\bar{a}_3,\bar{a}_5,\bar{a}_6,\xi)\sim (5\cdot10^5,10^6,10^5,5\cdot10^3,10^5),\nn
\end{align}
while tuning the abundance $f_{\rm PBH}$
\begin{align}
    \textrm{Toy model with an artificial dip} &\quad c(A_s,\sigma,\phi_0)\sim (10^5,10^5,5\cdot10^5)\nn \\
    \textrm{Minimally coupled polynomial} &\quad c(\bar{b}_2,\bar{b}_3,\bar{b}_5,\bar{b}_6)\sim (8\cdot10^8,2\cdot10^9,7\cdot10^8,7\cdot10^7)\nn \\
    \textrm{Non-Minimally coupled polynomial} &\quad c(\bar{a}_2,\bar{a}_3,\bar{a}_5,\bar{a}_6,\xi)\sim (8\cdot10^6,2\cdot10^7,10^6,7\cdot10^4,2\cdot10^6),\nn
\end{align}
which would lead to believe that a large degree of fine-tuning is present in these models ~\cite{Cole:2023wyx}, while in fact it signals a large amount of sensitivity. Indeed, as shown in Fig.~\ref{fig:fine-tuning}, independently of which parameters are varied, all models satisfy the naturalness criterion, namely $\gamma \simeq 1$, with only small and numerically irrelevant fluctuations.

While the criterion set by the parameter $c$  represents a practical realization of Wilson’s concept of naturalness, namely that observable features of a system should remain insensitive to infinitesimal changes in its fundamental parameters, it   captures only parameter sensitivity. The parameter $\gamma$, by properly normalizing 
the  naturalness measures,  is more appropriate to implement a variation of Wilson’s naturalness principle: observable properties should not exhibit anomalously large instabilities in response to small variations in the underlying parameters.

It is worth emphasising that the measure $\gamma$ is, by construction, a \emph{conditional} naturalness criterion: it answers the question of whether, \emph{given a specified physical regime for the observable $X$}, the parameter space exhibits a preferred direction or not. The integration range entering the average $\bar{c}$ is therefore not an arbitrary prior, but the precise specification of the physical question being asked. In our analysis the working assumption is the standard one in the PBH-as-dark-matter literature, namely that PBHs form with a cosmologically relevant abundance; the corresponding amplitude range follows from this assumption rather than being chosen to enforce $\gamma \sim \mathcal{O}(1)$. Within this regime, finding $\gamma \sim \mathcal{O}(1)$ for all the benchmark models considered is a non-trivial statement: it implies that no direction in parameter space is anomalously efficient at producing the required abundance. A different, complementary question, i.e. ``starting from an agnostic prior on $A$, is it natural for the model to enter the PBH-producing regime at all?'', is obtained by extending the integration range down to CMB-scale amplitudes $A \sim 10^{-9}$. In that enlarged domain the physics is no longer homogeneous, since most of the range corresponds to a regime where no USR phase is realised, and $\gamma$ correctly grows together with $c$, signalling the fine-tuning required to enter the PBH-producing regime. The two regimes thus address two distinct physical questions, both consistently captured by $\gamma$, while $c$ is by its local nature insensitive to this distinction.

\section{Conclusions}\label{sec:Conc}

In this work we have revisited the issue of naturalness in single-field inflationary models that generate PBHs through a phase of USR evolution. Rather than relying on qualitative fine-tuning arguments, we have introduced a quantitative and physically motivated measure of naturalness inspired by Wilsonian sensitivity criterion ~\cite{Anderson:1994dz,Anderson:1994tr,Anderson:1996ew}. 

While the criterion set by the parameter $c$  represents a practical realization of Wilson’s concept of naturalness, namely that observable features of a system should remain insensitive to infinitesimal changes in its fundamental parameters, it   captures only parameter sensitivity. The parameter $\gamma$, by properly normalizing 
the  naturalness measures,  is more appropriate to implement a variation of Wilson’s naturalness principle: observable properties should not exhibit anomalously large instabilities in response to small variations in the underlying parameters.

We have applied this criterion to three representative classes of inflationary potentials: a toy model with a localized feature, minimally coupled polynomial potentials, and polynomial potentials with a non-minimal coupling to gravity. For each case we have computed the scalar power spectrum numerically, verified consistency with CMB constraints on large scales and focused on scenarios producing PBHs in the asteroidal mass range, tuning the amplitude of the main peak in such a way to get a resulting PBH abundance in a wide range, from $f_{\rm PBH} \sim 10^{-7}$ up to order unity.

Our main result is that, for all models considered, the naturalness parameter $\gamma$ remains consistently of order unity. According to this criterion, PBH production in single-field inflationary scenarios are not technically unnatural.

We stress that this result should be read as a statement about \emph{abundance-level} naturalness, conditional on the working assumption that PBHs form with a cosmologically relevant abundance. A logically distinct question concerns the level of tuning required to enter the PBH-producing regime in the first place, i.e. to obtain a power-spectrum amplitude of order $10^{-2}$ starting from an agnostic prior that includes CMB-scale values $A \sim 10^{-9}$. Within our framework this question is recovered by extending the integration range entering $\bar{c}$, and yields a $\gamma$ that grows together with $c$, in line with the standard amplitude-level tuning of order $10^{4}$ reported in the single-field PBH literature\,\cite{Cole:2023wyx}. The fact that the local sensitivity $c$ does not distinguish between these two questions, while $\gamma$ does, is precisely the reason we adopt a normalised naturalness measure. Our claim that PBH-producing single-field models are not technically unnatural therefore refers to the abundance-level question and is fully compatible with the well-known amplitude-level tuning discussed in the literature.

A natural direction for future work is to extend this analysis to multi-field\,\cite{Garcia-Bellido:1996mdl,Bugaev:2011wy,Kawasaki:2015ppx,Clesse:2015wea,Braglia:2020eai,Palma:2020ejf,Braglia:2022phb,Balaji:2022dbi} and spectator\,\cite{Kawasaki:2012wr,Ando:2017veq,Ando:2018nge,Ferrante:2023bgz,Kawasaki:2021ycf,Chen:2019zza,Liu:2020zzv,Inomata:2020xad,Pi:2021dft,Gow:2023zzp,Inomata:2023drn,Lorenzoni:2025kwn} inflationary models of PBH production. In such scenarios, the enhancement of curvature perturbations can be generated or significantly modified after horizon exit, or even after inflation, potentially altering the small-scale power spectrum, the level of non-Gaussianity, and the resulting PBH mass function.

For instance, in Ref.~\cite{Stamou:2024lqf} the same sensitivity criterion, i.e. Eq.~\ref{a}, is employed as a measure of fine-tuning, and the author finds that in a specific spectator-field model one obtains $c \sim \mathcal{O}(1)$.
In particular, while the absolute sensitivity $c$ may be reduced in spectator-field models, $\gamma$ continues to highlight directions in parameter space that are atypical. For instance, in such models, the allowed range of the curvaton mass can be constrained by requiring $\gamma \sim \mathcal{O}(1)$, thereby providing a concrete physical interpretation of the naturalness measure.
Then, a systematic generalization of our analysis to include these additional ingredients, and to assess their impact on the naturalness measure $\gamma$, would provide a broader and more comprehensive understanding of the degree of naturalness of PBH formation models.

\section*{Acknowledgements}
We thank S. Allegrini, M. Braglia and F. Quevedo for useful comments and discussions. 
A.R.  acknowledges support from the  Swiss National Science Foundation (project number CRSII5\_213497).
\appendix
\setcounter{equation}{0}
\setcounter{section}{0}
\setcounter{table}{0}
\makeatletter
\renewcommand{\theequation}{A\arabic{equation}}

\bibliographystyle{JHEP}
\bibliography{main}

\end{document}